\documentclass[prb,twocolumn,aps,showpacs]{revtex4}
\usepackage{graphicx}
\def\ltsim{\vbox {\hbox{\lower .8\baselineskip \hbox{$<$}} \break
                 \hbox{\lower 0.2\baselineskip \hbox{$\sim$}} } }

\begin{document}

\title{Dephasing of mesoscopic interferences from  electron fractionalization}
\author{Karyn Le Hur}
\affiliation{D\'epartement de Physique, Universit\'e de Sherbrooke, 
Sherbrooke, Qu\'ebec, Canada J1K 2R1}

\begin{abstract}
We investigate the dephasing of mesoscopic interferences by electron-electron interactions in a strictly one-dimensional geometry composed of two weakly-coupled 
(clean and very long) Luttinger liquids.  The main goal of this paper is to demonstrate that in the present geometry interactions can produce a visible attenuation of Aharonov-Bohm oscillations. Through a Nyquist noise type description of the interactions 
and a direct (exact) calculation based on the Luttinger theory, in our setup we firmly emphasize that the dephasing time results from the electron fractionalization time.
\end{abstract}

\date{\today} 
\maketitle

Interesting phenomena in mesoscopic systems are known to result from quantum interferences: weak localization corrections to the conductivity, universal fluctuations of the conductance, and Aharonov-Bohm oscillations for example\cite{Imry}. The understanding of  ``dephasing" processes, {\it i.e.},  the physical causes which suppress those interference effects constitutes a topics of perpetual interest in mesoscopic systems. From the perspective of possible mesoscopic phase-coherent mesoscopic devices, knowledge of phase-breaking length or time is of great importance.   On the other hand, the loss of the electron phase coherence is interesting in its own right because this reveals information about the fundamental physics of the electron scattering mechanisms or electron decoherence in a correlated medium. An interfering particle  coupled to some environment fatally looses its phase. Notice that by environment we mean either some external dissipative bath\cite{cB,GM,MB} or still the electromagnetic field driven by the random thermal motion of other electrons in the system. It  has been indeed well-established that the effect of interactions in a disordered Fermi liquid can be embodied  by a fluctuating electromagnetic field or Nyquist noise\cite{bla}. At low temperatures, the predominant process generating dephasing in metals is irrefutably electron-electron interactions. In two dimensions (2D), experiments consistent with  the electron scattering time\cite{2D} $\tau_{\phi} \propto (T^2\ln T)^{-1}$ have been carried out in clean samples\cite{2Dexp}; $T$ being the temperature. 
Phase-breaking mechanisms in ballistic mesoscopic systems of dimensionality less than two are presently not completely understood and therefore would deserve some intensive theoretical and experimental endeavors. Recent Aharonov-Bohm oscillations measured on very clean (ballistic) quasi one-dimensional (1D) rings support a dephasing time which varies as\cite{Hansen} $\tau_{\phi}\propto T^{-1}$.  For quasi 1D disordered wires, one would rather expect\cite{bla,Imry} $\tau_{\phi}^{-1}\propto T^{2/3}$ as observed in Ref. \onlinecite{quasi1d}.
Of interest to us is to study dephasing in a non Fermi liquid system where electrons are not 
good quasiparticles. 

 \begin{figure}[htbp]
\begin{center}
\includegraphics[width=6cm,height=3.3cm]{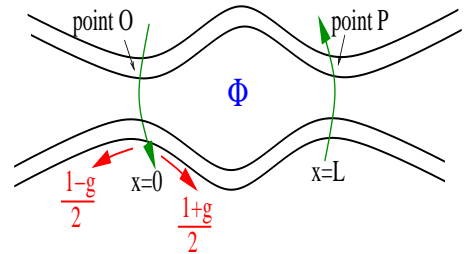}
\end{center}
\vskip -0.6cm
\caption{Electron  wave interferences with weakly-coupled Luttinger wires. The electrons can tunnel from one lead to the other at two point contacts located at $x=0$ and $x=L$.}
\label{setup}
\end{figure}

More precisely, the Luttinger liquid (LL) is well-known to exhibit fractional quasiparticles\cite{Glazman,Ines} which correspond to genuine excitations of one-dimensional ballistic systems with non-zero charge or/and current with respect to the ground state (with no plasmon excited)\cite{KV}. In Ref. \onlinecite{karyn}, exploiting the appropriate fractionalization mechanism, we have precisely derived the temperature and interaction dependence of the electron life-time in one dimension (1D). 
For spinless electrons, we report that the electron life-time obeys $\tau_{Fc}^{-1}\propto \pi T[(g+g^{-1})/2-1]$, $g<1$ being the well-known Luttinger exponent. For sufficiently weak interactions, in agreement with the recent Ref. \onlinecite{Mirlin}, we have found $\tau_{Fc}^{-1}\propto \pi T(Ua/\hbar v_F)^2$ 
where $U$ is the on-site interaction, $a$ the short-distance cutoff, and $v_F$ the Fermi velocity. 
On the other hand, after Ref. \onlinecite{karyn} it was still unclear which controlled experimental setup
could eventually detect this fractionalization time\cite{noteD}.
In this Letter, we propose to {\it unambiguously} reveal the latter via the dephasing of Aharonov-Bohm (AB) interferences built out from two weakly-coupled very long and spin-polarized Luttinger liquids  (which can be realized with quantum wires possessing a single conducting channel). First through two complementary approaches we show that in the geometry of Fig. 1 interactions inside the quantum wires can effectively suppress the AB oscillations, and second we firmly demonstrate that the resulting dephasing time $\tau_{\phi}$ can be identified as the electron fractionalization time $\tau_{Fc}$. 
As a matter of fact, a right-moving electron tunneling from one LL to the other, {\it e.g.}, at $x=0$, decomposes itself into a right-moving charge\cite{Ines,KV,karyn} $Q_+=(1+g)/2$ and a left-moving charge $Q_-=(1-g)/2$ inevitably provoking the loss of quantum interferences (in Refs. \onlinecite{KV,karyn}, $Q_{\pm}$ has been normalized to -e for convenience). 
For spinful electrons and weak interactions, the most important source of electron decoherence is spin-charge separation and the resulting electron life-time gets
modified as $\tau_{Fs}^{-1}=\tau_{\phi}^{-1}\propto TUa/(\hbar v_F)$. 

{\it Interactions as Nyquist noise}.--- It has been emphasized since more than a decade that an ohmic
environment could fake the electronic interactions in a one-channel mesoscopic conductor\cite{Nazarov}. In this picture, a one-channel conductor in series with a resistance is equivalent to a one-dimensional interacting system described by the LL. More precisely, the random thermal motion of electrons produces a fluctuating potential\cite{Girvin} $\delta V_i(t)$ which from the fluctuation-dissipation theorem is equivalent to an
effective resistance $R_j$ in each lead, $j=1$ being the upper lead and $j=2$ the lower lead; an
analogous point of view has been explored in the context of disordered Fermi
liquids\cite{bla}. Such a correspondence in 1D has been already 
established, {\it e.g.}, in the presence of a 
single-impurity\cite{IH} or a quantum dot\cite{KM} and we propose to extend it for this
setup of weakly-coupled LLs. 
In a given lead, a right-moving electron which is located at $x=0$ at the time $t=0$ will
propagate ballistically with a velocity $v_F$. Note that throughout the ohmic environment concept, at a frequency $\omega$ and for $0<x\leq L$, the related retarded electron Green's function takes a relatively simple form:
\begin{equation}
{\cal G}_{j}(x,\omega) = \exp \left[i\left(\frac{L\omega}{v_F} +m_j\pi\varphi\right)\frac{x}{L}+i{\cal K}_j\left(\frac{x}{v_F}\right)\right].
\end{equation}
The variable $x$ measures the position inside each wire.
The first term is the dynamical phase whereas the second term is the chirality-dependent AB phase
and the third term is induced by the fluctuating potential $\delta V_{j}(t)$ which is well-known to result  in an extra fluctuating phase $\pm {\cal K}_j(t)=\pm\frac{e}{\hbar}\int_0^t \delta V_{j}(t')dt'$ in the electron annihilation/creation operator of each lead. Here, $\varphi=\Phi/\Phi_o$ is related to the enclosed magnetic flux with $\Phi_o$
being the flux quantum, and the second term must exhibit a sign change ({\it e.g.}, $m_j=\pm$) for $j=1$ and $j=2$ respectively.

Let's consider an electron wave packet at the point O and examine the
interference phenomenon at the point P. The electron wave packet can take either the lead 1 or the lead
2 and the related transmission amplitudes are defined as ${\cal A}_1 = \sqrt{(1-T_0)(1-T_L)}{\cal G}_1(L,\omega)$ and ${\cal A}_2=\sqrt{T_0 T_{L}}{\cal G}_2(L,\omega)$; $T_i\ll 1$ with $i=0$ or $L$ denotes the tunneling probability at each point contact. Since $T_i\ll 1$, the transmission probability ${\cal T}=|{\cal A}_1+{\cal A}_2|^2$ from the point O to the point P hence can be approximated as:
\begin{equation}
{\cal T} \approx 1-(T_0+T_L)+\sqrt{T_0 T_{L}} \left(e^{i2\pi\varphi+iK_1-iK_2}+h.c.\right),
\end{equation}
$T_0\approx T_L$ and $K_j={\cal K}_j(\frac{L}{v_F})$; the crucial point being how the environments affect the partial waves. Similar to Refs. \onlinecite{MB,Nazarov,Girvin,IH}, we can resort to a set of harmonic oscillators to mimic the fluctuations of $\delta V_i$; this is especially well funded in 1D since interactions are known to produce bosonic type excitations (plasmons). Now, we can average ${\cal T}$ with respect  to the unperturbed set of oscillators exploiting the identity $\langle e^{iK_1-iK_2} \rangle = e^{-(\langle {K_1}^2+{K_2}^2\rangle)/2}$. Furthermore, the
variance $\langle {K_j}^2\rangle=\langle {\cal K}_j(\frac{L}{v_F})^2\rangle=({e^2}/{\hbar^2})\int_0^{L/v_F} dt' \int_0^{L/v_F} dt'' \langle \delta V_j(t') \delta V_j(t'') \rangle$ can be easily evaluated in the case of an ohmic environment (Nyquist noise). In the high-temperature regime, we must
identify\cite{MB} $\langle \delta V_j(t') \delta V_j(t'')\rangle = R_j \beta^{-1} \delta(t'-t'')$ which results in
$\langle {K_j}^2\rangle= 2\pi \beta^{-1} \frac{L}{\hbar v_F} r_j$ where $r_j=e^2 R_j/h$ represents the dimensionless resistance and $\beta=1/(k_B T)$. 

The final step attempts to relate the resistance $R_j$
with the corresponding Luttinger parameter $g_j$ which typically measures the strength
of the interactions in each wire. This can be performed by investigating the form of the local electron 
tunneling density of states
(TDOS) $\rho_j$ of each wire at one of the two contacts. From the environmental theory\cite{Nazarov,Girvin}, in which the interactions between electrons are embodied by a fluctuating potential, we extract\cite{note} $\rho_j(T)\propto T^{r_j}$ when $\beta^{-1}\ll \hbar v_F/a$. On the other hand, in the bulk of a LL, we have\cite{Glazman} $\rho_j(T)\propto T^{-1+(g_j+g_j^{-1})/2}$; this results in the precious equality $r_j=\frac{e^2 R_j}{h}=-1+(g_j+g_j^{-1})/2$ and then in
\begin{equation}
\langle {K_j}^2\rangle= 2\pi \beta^{-1} \frac{L}{\hbar v_F} \left(\frac{g_j+g_j^{-1}}{2}-1\right).
 \end{equation}
It is important to keep in mind that this formula is only appropriate in the weak-tunneling regime after
identification between the exact TDOS of a LL and that obtained from the environmental type theory. For $T_i\ll 1$, the average transmission probability $\langle {\cal T}\rangle$ exhibits the form:
\begin{equation}
\langle {\cal T} \rangle \approx \eta+\sqrt{4 T_0 T_L}\cos(2\pi\varphi) e^{\frac{-\pi L}{\hbar v_F \beta} \sum_{j} \left(\frac{g_j+g_j^{-1}}{2}-1\right)},
\end{equation}
where $\eta=1-(T_0+T_L)$. Applying the Landauer formalism, the total conductance of the wire 1 (similarly the conductance of the wire 2) then obeys
${\cal G}=(e^2/h)\langle {\cal T} \rangle$. The first term stems from $(e^2/h)|{\cal A}_1|^2$.  Moreover, as a blatant signature of dephasing due to electron-electron interactions the interference (flux-dependent) part of the conductance irrefutably exhibits an exponential suppression versus the temperature. The latter takes the specific form $\exp-2L/l_{\phi}$ where the dephasing length $l_{\phi}$ satisfies:
\begin{equation}
l_{\phi}^{-1}=\frac{\pi}{2\hbar v_F \beta} \sum_{j=1}^{j=2} \left(\frac{g_j+g_j^{-1}}{2}-1\right).
\end{equation}

Remember that in 1D, the dephasing length varies linearly with the thermal length $L_T\approx \hbar v_F \beta$. The visibility of the interference pattern will be suppressed when $L\geq l_{\phi}$ due to the fluctuations in the phases $K_1$ and $K_2$. We like to stress that the Nyquist noise description provides a relatively simple explanation of the result that in a 1D wire the dephasing length grows {\it linearly} with $L_T$; this stems from the fact that for an ohmic environment in the high-temperature limit the fluctuations of the potential $\delta V_j$ are proportional to the temperature\cite{GM}. This might be
relevant to explain the experiment of Ref. \onlinecite{Hansen} (however, their geometry is distinct from ours). For free electrons, implying $g_j=1$, we recover
 $l_{\phi}^{-1}=0$ and then perfect AB oscillations. It could be anticipated that like the suppression of the TDOS in 1D dephasing  may be attributed to electron fractionalization. Below, we like to enrich this Nyquist noise approach by a more direct (exact) calculation based on the Luttinger theory.

{\it Luttinger type calculation}.--- More precisely, the flux-dependent part of the current $\hbox{I}_{\varphi}$ in the wire 1 (or in the wire 2) passing between the points O and P as a function of the applied potential difference V may be calculated applying the Luttinger formalism for small tunneling amplitudes at the point contacts. The Hamiltonian $H=H_o+H_{tun}$ is the sum of the well-known Luttinger Hamiltonian $H_o$ as well as the tunneling part, $H_{tun}=\sum_{i}\Gamma_{i\pm}\Psi^{\dagger}_{2\pm}(x=i)\Psi_{1\pm}(x=i)+h.c.$, acting only at the point $i=0$ or $L$; the index $\pm$ refers to right and left movers respectively and for convenience we have denoted  $\Gamma_{0\pm}=\sqrt{T_0}\exp(i\mu_{1\pm}t/\hbar)$ and $\Gamma_{L\pm}=\sqrt{T_L}\exp(i\mu_{1\pm}t/\hbar)\exp(2i\pi\varphi)$. In contrast to the edges of quantum Hall systems, particles are not chiral, {\it i.e.}, they can propagate in both directions ``right'' or ``left'' and $\mu_{1\pm}$ refers to the electrochemical potential of each specy in the wire 1. We will
choose $\mu_{1+}=eV$ and $\mu_{1-}=0$;  another gauge would give the same physical result. 

To first order in $H_{tun}$, the current $\hbox{I}_{\varphi}(t)=-e\langle \dot{N}_{1+}\rangle$ where $N_{1+}$ is the number of right-moving electrons in the wire 1, takes the form $\hbox{I}_{\varphi}(t)=(ie/\hbar)\int dt' \theta(t-t') \hbox{Tr} \{\rho_o [\partial_t {N}_{1+}(t),{H}_{tun}(t')]\}$ where $\rho_o=e^{-\beta H_o}/\hbox{Tr} e^{-\beta H_o}$, in the interaction representation ${O}(t)=e^{i H_o t} O e^{-i H_o t}$, and $N_{1+}=\int dx\  \Psi^{\dagger}_{1+}(x)\Psi_{1+}(x)$. Again, this approach is appropriate to evaluate the
magnetic-flux dependent part of the current in the wire 1 because the latter can be treated perturbatively
in $\Gamma_{0+}$ and $\Gamma_{L+}$. Expressing $\partial_t N_{1+}(t)$ as  a function of $[N_{1+}(t),H_{tun}(t)]$ hence gives rise to $\hbox{I}_{\varphi}\propto -\sqrt{T_0 T_L}\left[e^{2i\pi\varphi}\Im m{X}_{L0}(\omega)+h.c.\right]_{\omega=eV/\hbar}$ where $X_{ij}(\omega)$ is 
the Fourier transform of ${X}_{ij}(t)=-i\theta(t)\langle[B_i(t),B^{\dagger}_j(0)]\rangle$ with $B_i=\Psi_{1+}(x=i)\Psi^{\dagger}_{2+}(x=i)$ and $i,j=0,L$ or vice-versa. 
The response function ${X}_{L0}(t)$ 
can be extracted resorting to standard bosonization techniques at finite temperature by simply analytically continuing\cite{Delft,Geller} $\tau\rightarrow it$ 
\begin{eqnarray}
X_{L0}(t) = \theta(t)\prod_{j=1}^2
\frac{1}{\sinh^{(\gamma_j+1)}\left[\frac{\pi}{L_{Tj}}(L-u_jt+im_j\delta)\right]} & &\\ \nonumber
\times \frac{a^{2\gamma_j}}{2\pi^2}\left(\frac{\pi}{L_{Tj}}\right)^{2\gamma_j+1}
\frac{1}{\sinh^{\gamma_j}\left[\frac{\pi}{L_{Tj}}(L+u_jt-im_j\delta)\right]}, & &
\end{eqnarray} 
where $u_j=v_F/g_j$ is  the plasmon velocity of each wire, the thermal length $L_{Tj}$ is precisely defined as $\hbar u_j \beta$, $\gamma_j=-1/2+(g_j+g_j^{-1})/4$, $\delta$ is a positive infinitesimal, and again $m_j=\pm$ for $j=1$ and $j=2$ respectively. We have implicitly considered the situation where $u_1\approx u_2=u$ assuming that the interaction strength between electrons is of the same order in magnitude in each wire; the relevant thermal length reads $L_T=\hbar u \beta$. 
The required Fourier transform may be calculated by contour integration and the poles are at $t=\frac{L}{u}\mp i\delta$ which asserts that the involved wave packets propagate at the plasmon velocity.  

Note that even though the previous environmental picture is not completely exact, {\it i.e.}, ignores this small renormalization effect of the electron velocity, this will only slightly renormalize the inverse of the dephasing length in Eq. (5) via an overall prefactor equal to $v_F/u$. The contributions from the two poles hence give rise to:
\begin{equation}
\hskip -0.09cm \hbox{I}{\varphi}\propto \frac{e^2 V}{h}\sqrt{T_0 T_L}\cos(2\pi\varphi)\frac{a^{2\gamma_1+2\gamma_2}[\pi/(L L_T)]^{\gamma_1+\gamma_2}}{\sinh^{\gamma_1} \frac{2\pi L}{L_T}\sinh^{\gamma_2} \frac{2\pi L}{L_T}};
\end{equation}
we have extracted the lowest order contribution in $V$ implying $V\rightarrow 0$. At relatively high temperatures $L\gg L_T$, the result can be approximated as $G_{\varphi}=d\hbox{I}_{\varphi}/dV \propto
(e^2/h)\sqrt{T_0 T_L}\cos(2\pi\varphi)a^{2\gamma_1+2\gamma_2}/[(L L_T)^{\gamma_1+\gamma_2}]
\exp-\left(\frac{2L}{L_{\phi}}\right)$; the {\it exact} dephasing length $L_{\phi}$ in 1D thus takes the form:
\begin{equation}
L_{\phi}^{-1}=\frac{\pi}{2\hbar u \beta} \sum_{j}  \left(\frac{g_j+g_j^{-1}}{2}-1\right)=l_{\phi}^{-1} \frac{v_F}{u}.
\end{equation}
First, it is important to bear in mind that the preceding Nyquist noise result is in a quite good agreement with the exact Luttinger type calculation. Second, it is also crucial to establish the clear physical origin of dephasing in 1D. Since the motion of
electrons is purely ballistic we can define the dephasing time as $\tau_{\phi}=L_{\phi}/u$. Recall that
for relatively weak interactions, equating $g_1=g_2=g$ and introducing the well-known formula 
$g=1-Ua/(\pi \hbar v_F)$ for the Luttinger parameter $g$ versus the Hubbard interaction $U$, we extract $\tau_{\phi}^{-1}\propto T(Ua/\hbar v_F)^2$. We note some resemblance with the dephasing times of Refs. \onlinecite{GM} and \onlinecite{Mirlin}. 

{\it $\tau_{\phi}$ as the electron fractionalization time}.--- At this step, it is certainly relevant to observe that $\tau_{\phi}=\hbar \beta/[-\pi+\pi(g+g^{-1})/2]$ is completely equivalent to the electron fractionalization time $\tau_{Fc}$ that we have built up in an earlier work\cite{karyn}. 
The fractionalization time $\tau_{Fc}$ has been precisely identified as follows. 
If one injects a right-moving electron in a 1D wire at the point $x=0$ at the time $t=0$, it is well-established\cite{KV,Ines} that this will fatally decompose into two counter-propagating modes, namely a charge $Q_+=(1+g)/2$  (normalized to -e) state going to the right at the plasmon velocity and a charge $Q_-=(1-g)/2$ state going to the left at the same velocity. Note that such a
fractionalization scheme reproduces nicely the properties (damping) of the exact electron Green's function\cite{karyn} and hence those of the TDOS. In Ref. \onlinecite{karyn} we have defined $\tau_{Fc}$ as the time needed for the propagator of the counter-going mode $Q_-$ to vanish at the position of the right-going mode $x\approx u\tau_{Fc}$;  at the time $\tau_{Fc}$, the overlap between the two fractional wave packets is negligible and the electron wave function gets clearly
 fractionalized. In the suggested geometry of Fig. 1,
an electron (a hole) which tunnels from the wire 1 (2) to the wire 2 (1) at $x=0$ gets subject to this fractionalization phenomenon producing the dephasing of electronic interferences. 
 It is essential that the wires are sufficiently long such that the reservoir leads attached at the
 extremities of each wire will not hinder the electron fractionalization mechanism emerging at $x=0$ (see Fig. 1); the length $d$ of each wire must satisfy $d\gg L$.

{\it Quantum limit}.--- So far we have only considered the relatively high-temperature limit $L\gg L_T=\hbar u \beta$. Now, we would like to briefly comment on the (opposite) quantum limit $\beta^{-1}\rightarrow 0$. Fom Eq. (7), we can easily extract $\hbox{I}_{\varphi} = (e^2/h)V\sqrt{T_0 T_L}\cos(2\pi\varphi)(a/L)^{2\gamma_1+2\gamma_2}$. Compared to the case of free electrons, one can notice an extra small power-law suppression as a function of the distance between the two point contacts. Nevertheless, if  $L$ is not too large compared to the lattice spacing $a$, one should observe visible electronic interferences when $T$ goes to zero. 
We may recover this result by applying the Nyquist environmental approach. In the case of an ohmic environment, in the quantum limit, it is easy to show that\cite{MB} $\langle {\cal K}_j(\frac{L}{v_F})^2\rangle \rightarrow -2r_j \ln(\omega_F L/v_F)$ where $\omega_F=v_F/a$. Now, using the important equality $r_j=2\gamma_j$ shown precedingly, $\hbox{I}_{\varphi}$ reaches that obtained above from the Luttinger theory.

{\it Electrons with spin}.--- One can extend the Luttinger theory developped above to the case
of electrons with spin. Here, the electron spectrum will exhibit both spin-charge separation and chiral decomposition from the charge sector\cite{KV,karyn}; the crucial point being that the spin 
propagates at the Fermi velocity $v_F$ whereas the fractional charge wave packets propagate at the charge plasmon velocity $u>v_F$. Hence this will produce four relevant poles at $t=(L/u)\pm i\delta$ and $t=(L/v_F)\pm i\delta$ which for {\it weak} interactions give the leading contribution 
\begin{eqnarray}
\hskip -0.1cm \hbox{I}{\varphi} &\propto & \frac{e^2 V}{h}\sqrt{T_0 T_L}\cos(2\pi\varphi) a^{\gamma_1+\gamma_2}L^2
 \frac{1}{\sinh^{\frac{\gamma_1}{2}} \left(\frac{2\pi L}{L_T}\right)} \\ \nonumber
&\times& \left(\frac{1}{L L_T}\right)^{1+\frac{\gamma_1+\gamma_2}{2}}
\frac{1}{\sinh^{\frac{\gamma_2}{2}} \left(\frac{2\pi L}{L_T}\right)\sinh\left[\frac{\pi}{L_T}(L-\frac{v_F}{u}L)\right]}. 
\end{eqnarray}
\vskip -0.2cm\hskip -0.4cm
Assuming $g_1=g_2=g$, the dephasing length obeys\cite{note2}
\begin{equation}
L_{\phi}^{-1}=\left[{\pi}/({2\hbar u \beta})\right]\left(\frac{g+g^{-1}}{2}-g\right),
\end{equation}
\vskip -0.2cm\hskip -0.4cm
and $\tau_{\phi}^{-1}\propto T(Ua/\hbar v_F)$; spin-charge separation accents dephasing compared to the spinless case. In Ref. \onlinecite{karyn},  when computing the fractionalization time $\tau_{Fs}$ for electrons with spin
we have omitted some relevant terms in\cite{Mirlin} ${\cal O}(u-v_F)$; when keeping those terms we check $\tau_{\phi}=\tau_{Fs}$. 

{\it Conclusion}.--- We have shed some light on the possibility to reveal the electron fractionalization mechanism occurring in 1D via a well-defined geometry composed of two {\it weakly-coupled} quantum wires. We have shown that the dephasing time related to the suppression of the Aharonov-Bohm oscillations at finite temperature is the electron fractionalization time.  We envision to extend this
work to different geometries and in particular to strongly-coupled quantum wires.
Finally Ref. \onlinecite{Heiblum} suggests to revisit dephasing for two coupled chiral LLs.

We are grateful to M. B\"{u}ttiker, I. Gornyi, A. Mirlin, and D. Polyakov for their comments on our Ref. \onlinecite{karyn}. This work was supported by CIAR, FQRNT, and NSERC.


\begin{thebibliography}{10}
\bibitem{noteD}
 A precise calculation of the current in a well-defined geometry has not been provided in Ref. \onlinecite{karyn}. It is still unclear to which realistic setup Eq. (14) could be applied.
\bibitem{note} When $T\rightarrow 0$, we find 
   $\langle {\cal K}_j(t)^2\rangle =-2r_j \ln(\omega_F t)$ and
   $\omega_F=v_F/a$. The Fourier transform of the electron Green's function gives
   $\rho_j(\omega)\propto \omega^{r_j}$ and $\rho_j(T)\propto T^{r_j}$ at finite T.
\bibitem{note2}
Note, the Nyquist approach would completely ignore the difference between spin and charge velocities and
would (incorrectly) give  $L_{\phi}^{-1}=\frac{\pi}{4L_T}(g+g^{-1}-2)\propto T(Ua/v_F)^2$.
\end{thebibliography}
\end{document}